\documentclass{article}
\usepackage[T1]{fontenc} 
\usepackage[utf8]{inputenc} 
\usepackage{ismir,amsmath,cite,url}
\usepackage{graphicx}
\usepackage{color}
\usepackage{textgreek}

\usepackage[bookmarks=false,hidelinks]{hyperref}

\usepackage{lineno}

\usepackage{xcolor}
\usepackage{amsmath}
\usepackage{amssymb}
\usepackage{amsthm}

\newcommand{\comment}[1]{}
\newcommand{\resolved}[1]{}

\newcommand{\G}{G}
\newcommand{\Dw}{D^\text{wave}}
\newcommand{\Ds}{D^\text{STFT}}
\newcommand{\Loss}{\mathcal{L}}
\newcommand{\Exp}{\mathbb{E}} 
\newcommand{\x}{x} 
\newcommand{\y}{y} 
\newcommand{\yh}{\hat{y}} 

\title{Learning to Denoise Historical Music}

\multauthor
{Yunpeng Li \hspace{1cm} Beat Gfeller \hspace{1cm} Marco Tagliasacchi \hspace{1cm} Dominik Roblek }%
{Google \\
{\tt\small \{yunpeng,beatg,mtagliasacchi,droblek\}@google.com}
}

\def\authorname{Y. Li, B. Gfeller, M. Tagliasacchi, and D. Roblek}

\begin{document}

\maketitle
\begin{abstract}
We propose an audio-to-audio neural network model that learns to denoise old music recordings. Our model internally converts its input into a time-frequency representation by means of a short-time Fourier transform (STFT), and processes the resulting complex spectrogram using a convolutional neural network. The network is trained with both reconstruction and adversarial objectives on a synthetic noisy music dataset, which is created by mixing clean music with real noise samples extracted from quiet segments of old recordings. We evaluate our method quantitatively on held-out test examples of the synthetic dataset, and qualitatively by human rating on samples of actual historical recordings. Our results show that the proposed method is effective in removing noise, while preserving the quality and details of the original music.
\end{abstract}

\section{Introduction}
\label{sec:intro}



Archives of historical music recordings are an important means for preserving cultural heritage.
Most such records, however, were created with outdated equipment, and stored on analog media such as phonograph records and wax cylinders.
The technological limitation of the recording process and the subsequent deterioration of the storage media inevitably left their marks, manifested by the characteristic crackling, clicking, and hissing noises that are typical in old records.
While ``remastering'' employed by the recording industry can substantially improve the sound quality, it is a time-consuming process of manual labor.
The focus of this paper is an automated method that learns from data to remove noise and restore music.

Audio denoising has a long history in signal processing~\cite{godsill1998}.
Traditional methods typically use a simplified statistical model of the noise, whose parameters are estimated from the noisy audio. Examples of these techniques are spectral noise subtraction~\cite{berouti1979, kamath2002}, spectral masking~\cite{reddy2007,grais2011},  statistical methods based on Wiener filtering~\cite{scalart1996} and Bayesian estimators~\cite{attias2001,loizou2005}.
Many of these approaches, however, focus on speech. Moreover, they often make simplifying assumptions about the structure of the noise, which makes them less effective on non-stationary real-world noise.

Recent advances in deep learning saw the emergence of data-driven methods that do not make such \emph{a priori} assumptions about noise.
Instead they learn an implicit noise model from training examples, which typically consist of pairs of clean and noisy versions of the same audio in a supervised setup.
Crucial challenges facing the adoption of the deep learning paradigm for our task 
are: i) can we design a model powerful enough for the complexity of music, yet simple and fast enough to be practical, and ii) how can we train such a model, given that we have no clean ground truth for historical recordings?
In this paper, we address these issues and show that it is indeed feasible to build an effective and efficient model for music denoising.



\subsection{Related Work}

%
Sparse linear regression with structured priors is used
 in~\cite{denoising-musical-audio-2008} to denoise music from synthetically added white Gaussian noise, obtaining large SNR improvements on a ``glockenspiel'' excerpt, and on an Indian polyphonic song.
%
\cite{DengSESZFO20} considers the problem of removing artifacts of perceptual coding audio compression with low bit-rates. That work, which uses LSTMs, is the first successful application of deep learning for this type of music audio restoration.
Note that in contrast to our work, aligned pairs of original and compressed audio samples are readily available.
Statistical methods are applied in~\cite{greek-folk-music-denoise-2014} to denoise Greek Folk music recorded in outdoor festivities.
In \cite{denoising-phonogram-2015}, the author applies structured sparsity models to two specific audio recordings that were digitized from wax cylinders, and describes the results qualitatively.
In~\cite{inpainting-long-audio-segments-2018}, the authors describe how to fill in gaps (at known positions) of several seconds in music audio, using self-similar parts from the recording itself.

Our method is also related to audio super-resolution, also known as bandwidth extension. This is the process of extending audio from low to higher sample rates, which requires
restoring the high frequency content. In~\cite{audio-superresolution-2017,temporal-film-2019} two approaches which work for music are described. On piano music, for example, \cite{temporal-film-2019} obtains an SNR of 19.3 when up-sampling a low-pass filtered audio from 4kHz to 16kHz.

Many existing denoising approaches focus on speech instead of music~\cite{deep-network-priors-2019,germain2018senet,segan-icassp-2017,rethage2018}. Given that these two domains have
very different properties, it is not clear a priori how well such methods transfer to the music domain. Nevertheless, our work is inspired by recent approaches that use generative adversarial networks (GANs) to improve the quality of audio~\cite{segan-icassp-2017,donahue2017,audio-codec-icassp2020}. For example,~\cite{audio-codec-icassp2020} obtains significant improvements denoising speech and applause sounds that have been decoded at a low bit-rate, using a wave-to-wave convolutional architecture. 










In this paper, we present a method to remove noise from historical music recordings, using two
sources of audio: i) a collection of historical music recordings to be restored, for which no clean reference is available, and ii) a separate collection of music of the same genre that contains high-quality recordings. We focus on classical music, for which both public domain historical recordings as well as modern digital recordings are available.
This paper makes the following contributions:
\begin{itemize}
    \item We provide a fully automated approach that succeeds in removing noise from historical recordings, while preserving the musical content in high quality. Quality is measured in terms of SNR and subjective scores inspired by MUSHRA~\cite{mushra-2015},
    and examples on real historical recordings are provided\footnote{\url{https://www.youtube.com/playlist?list=PLa5CkN3odpnxi3WqMH4MgVk7XUjCP99d3}}.
    
    \item Our approach employs a new architecture that transforms audio in the time domain, using a multi-scale approach, combined with STFT and inverse STFT. As this architecture is able to output high-quality music, it may be a useful architecture for other tasks that involve the transformation of music audio.
    \item We provide an efficient and fully automated method to extract noise segments (without music) from a collection of historical music recordings. This is a key ingredient of our approach, as it allows us to create synthetic pairs of <clean, noisy> audio samples.
\end{itemize}

The rest of this paper is organized as follows.
Our approach is described in detail in Section~\ref{sec:method}, and experimental results are given in Section~\ref{sec:experiments}. We conclude in Section~\ref{sec:conclusion}.

\section{Method}
\label{sec:method}

Our model is an audio-to-audio generator learned from paired examples with both reconstruction and adversarial objectives.


 
\subsection{Creating paired training examples}
\label{sec:noise-extraction}
For training, we use time-aligned pairs of <clean, noisy> examples, where clean music is used as targets, and noisy music as inputs to the generator. We take a data-driven approach to generate noisy audio from clean references. We synthesize noisy samples by simulating the degradation process affecting the historical recordings, namely applying band-pass filtering, followed by additive mixing with noise samples extracted from ``quasi-silence'' segments of historical recordings.

 Specifically, we scan the noisy historical recordings looking for low-energy segments in the time domain, which corresponds to pauses in the musical scores. To this end, we compute the rolling standard deviation from the raw audio samples with a window size equal to 100ms. Then, we estimate an adaptive threshold $\tau$ based on the $q$-th quantile of the standard deviations and keep the segments that satisfy the following two conditions: i) the local standard deviation is below $\tau$, and ii) the segment has a minimum duration of $T$. Intuitively, the value of $q$ is selected based on a trade-off between the number of extracted segments and the need of extracting noise-only segments. In our experiments, we set $q = 0.5$\% and $T=100$ms. In this way, from 801 different recordings, we are able to extract around 8900 noise samples.

From each of these short noise segments, we need to generate noise samples having the same length as the clean audio references. We do this by replicating the noise segment in time, using overlap-and-add (OLA) with an overlap equal to 20\% of the segment length. Given the short duration of most noise segments, this operation alone would lead to periodic noise patterns which differ from the noise characteristics found in historical recordings. Therefore, we alter each noise segment replica before the OLA synthesis step in two ways: i) applying a random perturbation to the phase of the noise segment (adding Gaussian noise $\sim\mathcal{N}(0, 0.1)$ to the phase of the STFT); ii) applying a random shift in time (with wraparound). We found that these simple operations produce longer noise samples with auditory characteristics similar to the ones encountered in the historical recordings, avoiding artificial periodic patterns. 

Finally, we create time-aligned pairs of <clean, noise> examples by: i) applying band-pass filtering with cutoff frequencies randomly sampled in [50Hz, 150Hz] and [5kHz, 10kHz], respectively; ii) mixing a randomly selected noise sample with a gain in the range [10dB, 30dB].


\subsection{Model architecture}


The generator processes the audio in the time-frequency domain. 
It first computes the STFT of the input, the real and imaginary components of which are then fed as a 2-channel image to a 2D convolutional U-Net~\cite{unet} followed by an inverse STFT back to the time domain. Finally the output is added back to the input, making the model a residual generator.

\begin{figure}[t]
\centering
\includegraphics[width=0.47\textwidth]{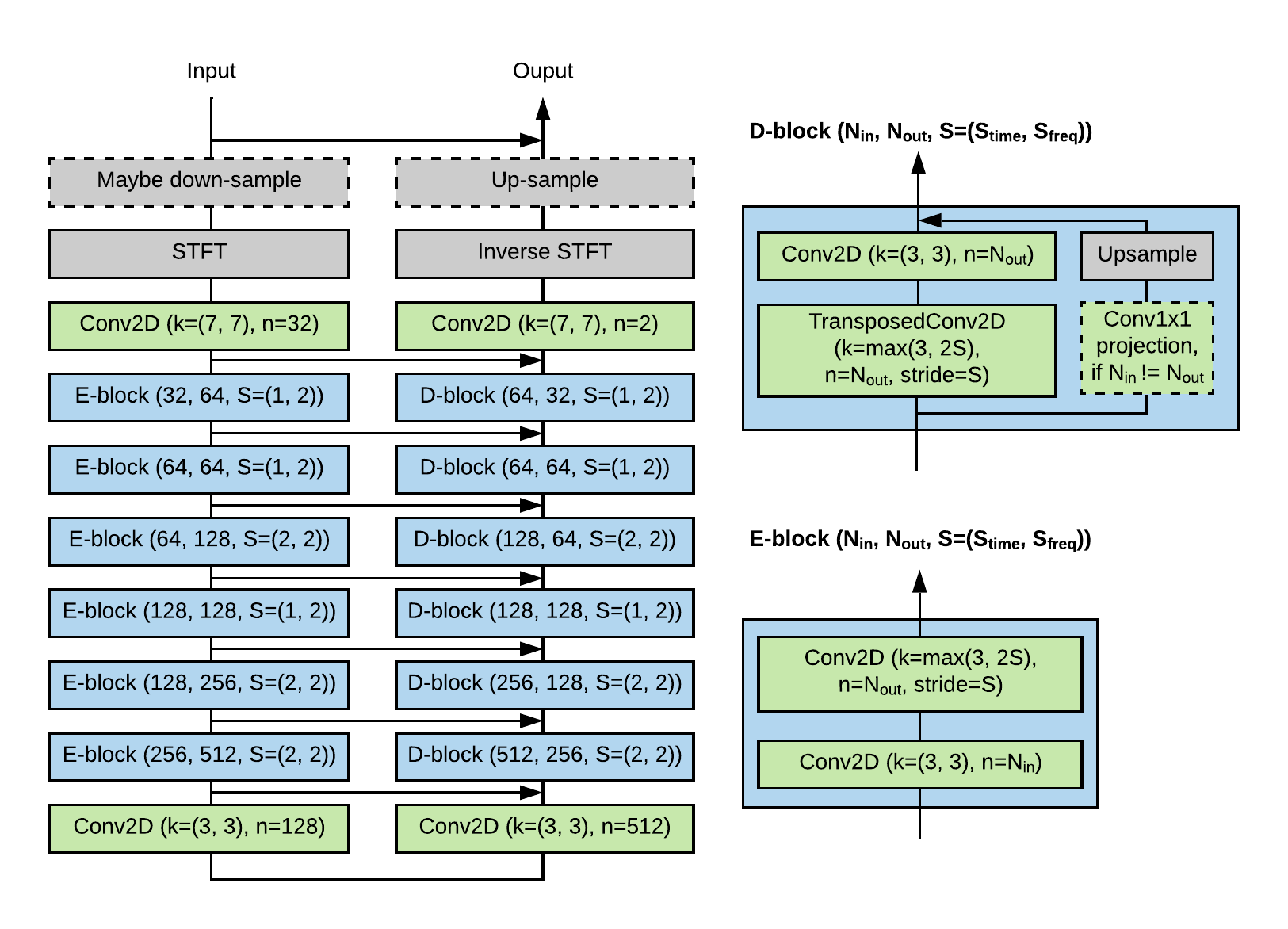}
\caption{
    Generator architecture. 
    Dashed-line components are included on a need-to-have basis: Up/down-sampling of the input/output audio is needed for processing at coarser resolutions in a multi-scale setup; The linear projection (by 1x1 convolution) in the decoder block is present only when the output of the block has a different number of channels from its input.}
\label{fig:generator_arch}
\end{figure}


The U-Net in our generator is a symmetric encoder-decoder network with skip-connections, where the architecture of the decoder layers mirrors that of the encoder and the skip-connections run between each encoder block and its mirrored decoder block.
Each encoder block is a 3$\times$3 convolution followed by either a 3$\times$4 convolution with stride of 1$\times$2 (if down-sampling in the frequency dimension), or a 4$\times$4 convolution with stride of 2$\times$2 (if down-sampling in both time and frequency dimensions).
We choose kernel sizes to be multiples of strides to ensure even contribution from all locations of the input feature map, which prevents the formation of checkerboard-like patterns in resampling layers~\cite{odena2016deconvolution}.
The decoder blocks mirror the encoder blocks, and each consists of a transposed convolution for up-sampling followed by a 3$\times$3 convolution.
Each decoder block additionally includes a shortcut connection between its input and output. The shortcut consists of a nearest-neighbor up-sampling layer, which is followed by a linear projection using 1x1 convolution when the output has a different number of channels from the input.
We do not include a shortcut in the encoder block, since it already shares the same input with a U-Net skip connection and therefore only needs to produce the residual complementary to the skip path.
The architecture of the generator is shown in Figure~\ref{fig:generator_arch}.

We use two discriminators for the adversarial objective, one in the waveform domain and one in the STFT domain.
The STFT discriminator has the same architecture as the encoder module of the generator. For the waveform discriminator, we use the same architecture as MelGAN~\cite{kumar2019} except that we only double (instead of quadruple) the number of channels in the down-sampling layers. We found this light-weight version to be sufficient in our setup, and that using the full version had no additional benefit.
Both discriminators are fully convolutional. Hence the waveform discriminator produces a 1D output spanning the time domain, and the STFT discriminator has a 2D output spanning the time-frequency domain.

We use weight normalization~\cite{salimans2016} and ELU activation~\cite{clevert2016} in the generator, while layer normalization~\cite{ba2016} and Leaky ReLU activation~\cite{maas2013} with $\alpha=0.3$ are used in the discriminator.

\subsubsection{STFT Representation}
In the generator, the STFT is represented by a 2-channel image, where the channels are the real and imaginary components.
We also explored a polar representation, where the channels are the modulus and the phase; additionally we experimented with processing only the modulus channel and reusing the original phase, as is done in~\cite{abdulatif2019aegan}. Nevertheless, we found the real/imaginary representation to perform better in our experiments.

Furthermore, we tried aligning the phase so that the phase in each frame is coherent with a global reference (e.g., the first frame) rather than its local STFT window. Again, we observed no advantage in doing so, which suggests that the neural network is capable of internally handling the phase offsets.
Unlike~\cite{abdulatif2019aegan}, we do not convert STFT to logarithmic scale as we found it be detrimental to performance (even with various smoothing and normalization schemes).

\subsubsection{Multi-scale Generator}
We can further stack multiple copies of the generator described above, each with its own separate parameters, in a coarse-to-fine fashion:
The generators at earlier stages process the audio at reduced temporal resolutions, 
whereas the later-stage generators focus on restoring finer details. This is equivalent to halving the sampling rate in each scale.
This type of multi-scale generation scheme is routinely used in computer vision and graphics to produce high-resolution images
(e.g.,~\cite{karras2018progressive-gan}).

Let $K$ be the total number of scales, then generator $G_k$ at scale $k$ ($k \in \{0, \dots, K-1\})$ down-samples its input by a factor of $2^k$ before computing the STFT and up-samples the output residual (after computing the inverse STFT) by the same factor to match the resolution of the input.
The overall generator $G$ is the composite of $G_0 \circ \dots \circ G_{K-1}$.


Compared with simply stacking U-Nets all  at the original input resolution, as done in~\cite{armanious2018medgan}, the benefit of the multi-scale approach is two-fold: i) the asymptotic computational complexity is constant with respect to the number of scales, as opposed to linear in~\cite{armanious2018medgan}, due to exponentially decreasing input sizes at coarser levels; ii) the intermediate outputs of the generator correspond to the input audio processed at lower resolutions, which allows us to meaningfully impose multi-scale losses on the intermediate outputs in addition to the final output. We will describe how this can be accomplished in the next section.

\subsection{Training}
\label{subsec:training}

The generator can be trained using the reconstruction loss between the denoised output and the clean target. This can be further complemented with an adversarial loss, given by discriminators trained simultaneously with the generator, a practice often used in audio enhancement (e.g.,~\cite{segan-icassp-2017,donahue2017,abdulatif2019aegan}, among others). 
%
%
In the case of our multi-scale generator, we use the same number of waveform and STFT discriminators as generator scales.
This way, there is one discriminator of both types for each of the (down-sampled) intermediate outputs and final output in each domain.
For the adversarial loss, we use the hinge loss averaged over multiple scales.
Since the discriminators are convolutional, this loss is further averaged over time for the waveform discriminator and over time-frequency bins for the STFT discriminator. Similarly, the reconstruction loss is also imposed on the outputs at each scale.

More formally, let $(\x, \y)$ denote a training example, where $\x$ is the noisy input and $\y$ is the clean target, and $k \in \{0, \dots, K-1\}$ denote the scale index.
Hence $\y_k$ is the clean audio down-sampled to scale $k$, and $\yh_k$ represents the intermediate output of the generator $G_k \circ \dots \circ G_{K-1}(\x)$ down-sampled to the same scale. Note that for the finest scale $k=0$ at full resolution, $\y_0=\y$ is simply the original clean audio and $\yh \triangleq \yh_0 = G(\x)$ is the final output of the generator.
Thus the $L^1$ reconstruction loss in the STFT domain can be written as
\begin{equation}
    \Loss_G^{\text{rec}} = \Exp_{(x,y)} \left[ \sum_{k} \frac{\| \omega_k - \hat{\omega}_k \|_1}{S^\text{STFT}_k} \right],
\end{equation}
where 2D complex tensors $\omega_k$ and $\hat{\omega}_k$ denote the STFT of down-sampled clean audio $\y_k$ and generator output $\yh_k$ for scale $k$, respectively, and $S^\text{STFT}_k$ is the total number of time-frequency bins in $\omega_k$ and $\hat{\omega}_k$.
We find this STFT-based reconstruction loss to perform better than either imposing per-sample losses directly in the waveform domain or using losses computed from the internal ``feature'' layers of discriminators (e.g.~\cite{kumar2019}).

For the adversarial loss, let $t$ denote the temporal index over all $T_k$ logits of the waveform discriminator at scale $k$ (recalling that the discriminators are fully convolutional) and let $s$ denote the index over all $S_k$ logits of the STFT discriminator. 
Then discriminator losses in the wave and STFT domains can be written as, respectively,
\begin{align}
    \nonumber
    \Loss_\Dw =~&\Exp_{\y} \left[\sum_{k,t} \frac{1}{T_k} \max(0, 1 - \Dw_{k,t}(\y_k)) \right] + \\  
    &\Exp_{\x} \left [ \sum_{k,t} \frac{1}{T_k} \max(0, 1 + \Dw_{k,t}(\yh_k)) \right ] \\
    \nonumber
    \Loss_\Ds =~&\Exp_{\y} \left[\sum_{k,s} \frac{1}{S_k} \max(0, 1 - \Ds_{k,s}(\y_k)) \right] + \\ 
    &\Exp_{\x} \left [ \sum_{k,s} \frac{1}{S_k} \max(0, 1 + \Ds_{k,s}(\yh_k)) \right ],
\end{align}
and the corresponding adversarial loss for the generator is given by
\begin{align}
    \nonumber
    \Loss^\text{adv}_\G =~ & \Loss^\text{adv, wave}_\G + \Loss^\text{adv, STFT}_\G \\ \nonumber
    =~  & \Exp_{\x} \left[ \sum_{k,t} \frac{1}{T_k} \max(0, 1 - \Dw_{k,t}(\yh_k)) \right. + \\
        & \qquad \left. \sum_{k,s} \frac{1}{S_k} \max(0, 1 - \Ds_{k,s}(\yh_k)) \right].
\end{align}
%

The overall generator loss is a weighted sum of the adversarial loss and the reconstruction loss, i.e.,
\begin{equation}
    \Loss_\G = \Loss_G^{\text{rec}} + \lambda \cdot \Loss_G^{\text{adv}}.
\end{equation}

We set the weight of the adversarial loss $\lambda$ to $0.01$ in all our experiments, except those where we do not use discriminators (which corresponds $\lambda$=0). 
We train the model with TensorFlow for 400,000 steps using the ADAM~\cite{adam-optimizer} optimizer, with a batch size of 16 and a constant learning rate of 0.0001 with $\beta_1=0.5$ and $\beta_2=0.9$.
For the STFT, we use a window size of 2048 and a hop size of 512 when there is only a single scale.
For each added scale we halve the STFT window size and hop size \emph{everywhere}. This way the STFT window at the coarsest scale has a receptive field of 2048 samples at the original resolution, whereas finer levels have smaller receptive fields and hence focus more on higher frequencies.

Our model has around 9 million parameters per scale in the generator. At inference-time, it takes less than half a second for every second of input audio on a modern CPU and more than an order of magnitude faster on GPUs.
\section{Experiments}
\label{sec:experiments}

We evaluate our model on a dataset of synthetically generated noisy-clean pairs, using both objective and subjective metrics. In addition, we also provide a subjective evaluation on samples from real historical recordings, for which the clean references are not available. 

\subsection{Datasets}
\label{subsec:datasets}

Our data is derived from two sources: i) digitized historical music recordings from the Public Domain Project~\cite{public-domain-project-histrec}, and ii) a collection of classical music recordings of CD-quality.
The historical recordings are used in two ways: i) to extract realistic noise from relatively silent portions of the audio, as described in Section~\ref{sec:noise-extraction}; and ii) to evaluate different methods based on the human-perceived subjective quality of their outputs.
The modern recordings are used for mixing with the extracted noise samples to create synthetic noisy music, as well as serving as the clean ground truth.
We additionally filter our data to retain only classical music, as it is by far the most represented genre in historical recordings.
%
%
The resulting dataset consists of pairs of clean and noisy audio clips, both monophonic and 5 seconds long, sampled at 44.1kHz. The total duration of the clean clips is 460h. 


\subsection{Quantitative Evaluation}
\label{subsec:quantitative-eval}

We quantitatively evaluate the performance of different methods on a held-out test set of 1296 examples from the synthetic noisy music dataset.
For the neural network models, whose training is stochastic, we repeat the training process 10 times for each model and report the mean for each metric and its standard error.

\textbf{Evaluation metrics:} Objective metrics such as the signal-to-noise ratio (SNR) faithfully measure the difference between two waveforms on a per-sample basis, but they often do not correlate well with human-perceived reconstruction quality.
Therefore, we additionally measure the \emph{VGG distance} between the ground truth and the denoised output, which is defined as the $L^2$ distance between their respective embeddings computed by a \mbox{VGGish} network~\cite{hershey2017vggish}. The embedding network is pre-trained for multi-label classification tasks on the YouTube-100M dataset, in which labels are assigned automatically based on
a combination of metadata (title, description, comments, etc.), context, and image content for each video. Hence we expect the VGG distance to focus more on higher-level features of the audio and less on per-sample alignment.
Note that the same embedding used by Frech\'{e}t audio distance (FAD)~\cite{fad-interspeech-2019}, which measures the distance between two \emph{distributions}. However, FAD does not compare the content of individual audio samples, and is hence not applicable to denoising.


\newcommand{\ms}[2]{{#1}\footnotesize{$\pm${#2}}}

\begin{table}[t]
    \centering
    \begin{tabular}{|l|c|c|}
        \hline
         & $\Delta$SNR (dB) & -$\Delta$VGG \\
        \hline
        1 scale & \textbf{\ms{3.4}{0.0}} & \ms{0.68}{0.01} \\
        2 scales & \textbf{\ms{3.4}{0.0}} & \textbf{\ms{0.78}{0.01}} \\ 
        3 scales & \ms{3.2}{0.0} & \ms{0.73}{0.01} \\  
        \hline
    \end{tabular}
    \caption{Performance of our model with different numbers of scales $K$ in terms of SNR gain ($\Delta$SNR) and VGG distance reduction (-$\Delta$VGG). Higher is     better.}
    \label{tab:quantitative-scales}
\end{table}

We report the SNR gain ($\Delta$SNR) and VGG distance reduction (-$\Delta$VGG) of the denoised output relative to the noisy input, averaged over the test set.
For reference, the noisy input has an average SNR of 14.4dB and VGG distance of 2.09.
Table~\ref{tab:quantitative-scales} shows the performance of our model with different numbers of scales.
We use $K=2$ scales for the rest of our experiments.
\begin{table*}[ht]
    \centering
    \begin{tabular}{|l|c|c|c|c|c|c|c|c|}
        \hline
         & \multicolumn{4}{c|}{$\Delta$SNR (dB)} & \multicolumn{4}{c|}{-$\Delta$VGG} \\
         & \multicolumn{3}{c|}{noise level} & & \multicolumn{3}{c|}{noise level} &    \\
         & low & medium & high & all & low & medium & high & all \\
        \hline
        Ours, $\lambda$=0 & \textbf{\ms{2.5}{0.0}} & \textbf{\ms{4.1}{0.0}} & \textbf{\ms{4.3}{0.0}} & \textbf{\ms{3.7}{0.0}} & \ms{0.30}{0.01} & \ms{0.47}{0.01} & \ms{0.58}{0.01} & \ms{0.45}{0.01} \\
        Ours, $\lambda$=0.01 & \ms{2.2}{0.0} & \ms{3.9}{0.0} & \ms{4.1}{0.0} & \ms{3.4}{0.0} & \textbf{\ms{0.66}{0.01}} & \textbf{\ms{0.81}{0.01}} & \textbf{\ms{0.87}{0.01}} & \textbf{\ms{0.78}{0.01}} \\
        Ours, bypass phase & \ms{2.1}{0.0} & \ms{3.5}{0.0} & \ms{3.7}{0.0} & \ms{3.1}{0.0} & \ms{0.62}{0.01} & \ms{0.77}{0.01} & \ms{0.83}{0.01} & \ms{0.74}{0.01} \\
        MelGAN-UNet & \ms{1.7}{0.0} & \ms{2.9}{0.0} & \ms{3.1}{0.0} & \ms{2.6}{0.0} & \ms{0.16}{0.02} & \ms{0.15}{0.03} & \ms{0.18}{0.02} & \ms{0.16}{0.02} \\
        DeepFeature generator & \ms{-0.7}{0.4} & \ms{1.3}{0.1} & \ms{1.7}{0.1} & \ms{0.8}{0.2} & \ms{0.00}{0.02} & \ms{0.03}{0.02} & \ms{0.00}{0.01} & \ms{0.01}{0.02} \\
        \hline
        log-MMSE & -1.4 & -0.2 & 0.1 & -0.5 & -0.15 & -0.04 & 0.01 & -0.07 \\
        Wiener & 0.1 & 0.1 & 0.1 & 0.1 & 0.01 & 0.02 & 0.01 & 0.01 \\
        \hline
    \end{tabular}
    \caption{Performance of different variants of our model and alternative approaches, evaluated on subsets of examples with different noise levels as well as on     the full test set.}
    \label{tab:quantitative-all}
\end{table*}
We evaluate variants of our proposed model in an ablation study and compare with alternative approaches and well-established signal processing baselines:
\begin{itemize}
    \item \textbf{Ours, $\lambda$=0}: Our model trained with only reconstruction loss.
    \item \textbf{Ours, $\lambda$=0.01}: Our model trained with both adversarial and reconstruction losses.
    \item \textbf{Ours, bypass phase}: Same as above, except that the phase of the noisy input is reused and only the modulus of the STFT is processed by the U-Net (as a single-channel image). This is similar to the approach of~\cite{abdulatif2019aegan}, but trained and evaluated for music denoising instead of speech.
    \item \textbf{MelGAN-UNet}: A 1D-convolutional waveform-domain generator inspired by MelGAN~\cite{kumar2019}, where the decoder is the same as the generator of MelGAN and the encoder mirrors the decoder.
    \item \textbf{DeepFeature generator}: The 1D-convolutional waveform-domain generator of~\cite{germain2018senet}, which does not use U-Net but rather a series of 1D convolutions with exponentially increasing dilation sizes. Unlike U-Net, the temporal resolution and number of channels remain unchanged in all layers of this network.
    \item \textbf{log-MMSE}: A short-time spectral amplitude estimator for speech signals which minimizes the mean-square error of the log-spectra~\cite{ephraim1985}. In our implementation, the estimation of the noise spectrum is based on low-energy frames across the whole clip, rather than considering the frames at the start of the audio clip. We use this deviation from the standard implementation as it gives better SNR results.
    \item \textbf{Wiener}: A linear time-invariant filter that minimizes the mean-square error. We adopted the SciPy~\cite{scipy2020} implementation and used default parameters, as 
    different parameter settings did not improve the results.
\end{itemize}
For waveform-domain generators, we tried waveform-domain losses -- including reconstruction losses in the ``feature space'' of discriminator internal layers~\cite{germain2018senet,kumar2019} -- as well as STFT-domain losses, and found the former to work better with the DeepFeature generator while the latter gave better results for the MelGAN-UNet generator.
The results shown for these generators are those obtained with the better loss variant.
We also divide the test set into three subsets, each containing the same number of examples, with low noise (avg. 19.8dB SNR), medium noise (avg. 14.2dB SNR), and high noise (avg. 9.4dB SNR), and compute the same metrics on each subset as well as on the full test set. 

The results in Table~\ref{tab:quantitative-all} show that, for all noise levels, our model consistently outperforms the signal processing baselines and the waveform-domain neural network models, which have proven highly successful in speech enhancement but are not adequate for the complexity of music signals.
The signal-processing baselines (log-MMSE and Wiener filtering) are hardly able to improve upon the noisy input at all. This is not too surprising given the non-Gaussian, non-white nature of the real-world noise in the evaluation data.
Comparing the results among the variants of our model, we further make the following observations:
\begin{itemize}
    \item Using adversarial losses does not help in terms of SNR, as is evident from the top two rows of Table~\ref{tab:quantitative-all}. The SNR decrease is small but significant. The adversarially trained variant, however, scores better on the high-level feature oriented VGG distance metric, which is in line with past observations~\cite{segan-icassp-2017,kumar2019}
    \item It is advantageous to take both the modulus and the phase into account when processing the STFT spectrogram, as the ``bypass-phase'' variant which reuses the input phase produces consistently worse results across all noise levels. This shows that the proposed model is able to reconstruct the fine-grained phase component of the original clean music.
\end{itemize}

\subsection{Subjective Evaluation}
In the previous section we compared results by means of objective quality metrics, which can be quantitatively computed from pairs of noisy-clean examples. These metrics can be conveniently used to systematically run an evaluation over a large number of samples. However, it is difficult to come up  with an objective metric that correlates with quality as perceived by human listeners.
Indeed, the SNR and VGG distance metrics do not agree in our quantitative evaluation -- the proposed model is better in terms of VGG distance, but worse in terms of SNR compared to its counterpart without discriminator. We now describe our subjective evaluation which we ran in order to identify the method that performs best when judged by humans.

Following recent work on low-bitrate audio improvement~\cite{audio-codec-icassp2020},
we use a score inspired by MUSHRA~\cite{mushra-2015} for our subjective evaluation.
Each rater assigned a score between 0 and 100 to each sample.
The main difference to actual MUSHRA scores is that since no clean reference exists for historical recordings, we do not include an explicit reference in the rated samples (although we do include the clean sample in the synthetic dataset evaluation).

We perform our evaluation on 10 samples of historical recordings, and separately on 10 samples from the synthetic dataset, using 11 human raters. As in the objective evaluation, each sample is 5 seconds long.
We evaluate the following four versions for each sample:
(i) Original historic audio example, (ii) denoised example using our model with $\lambda$=0.01, (iii) denoised example using our model with $\lambda$=0,
(iv) denoised example using log-MMSE.

For the synthetic dataset, we use the four versions above, but instead of the historic audio we use the synthetically noisified one.
We do not include Wiener filtering as a competing baseline here since we noticed that it produces outputs that are consistently near-identical to the noisy input, and hence including it in the subjective evaluation would provide little value.
\begin{figure}[t]
\centering
\includegraphics[width=0.95\linewidth]{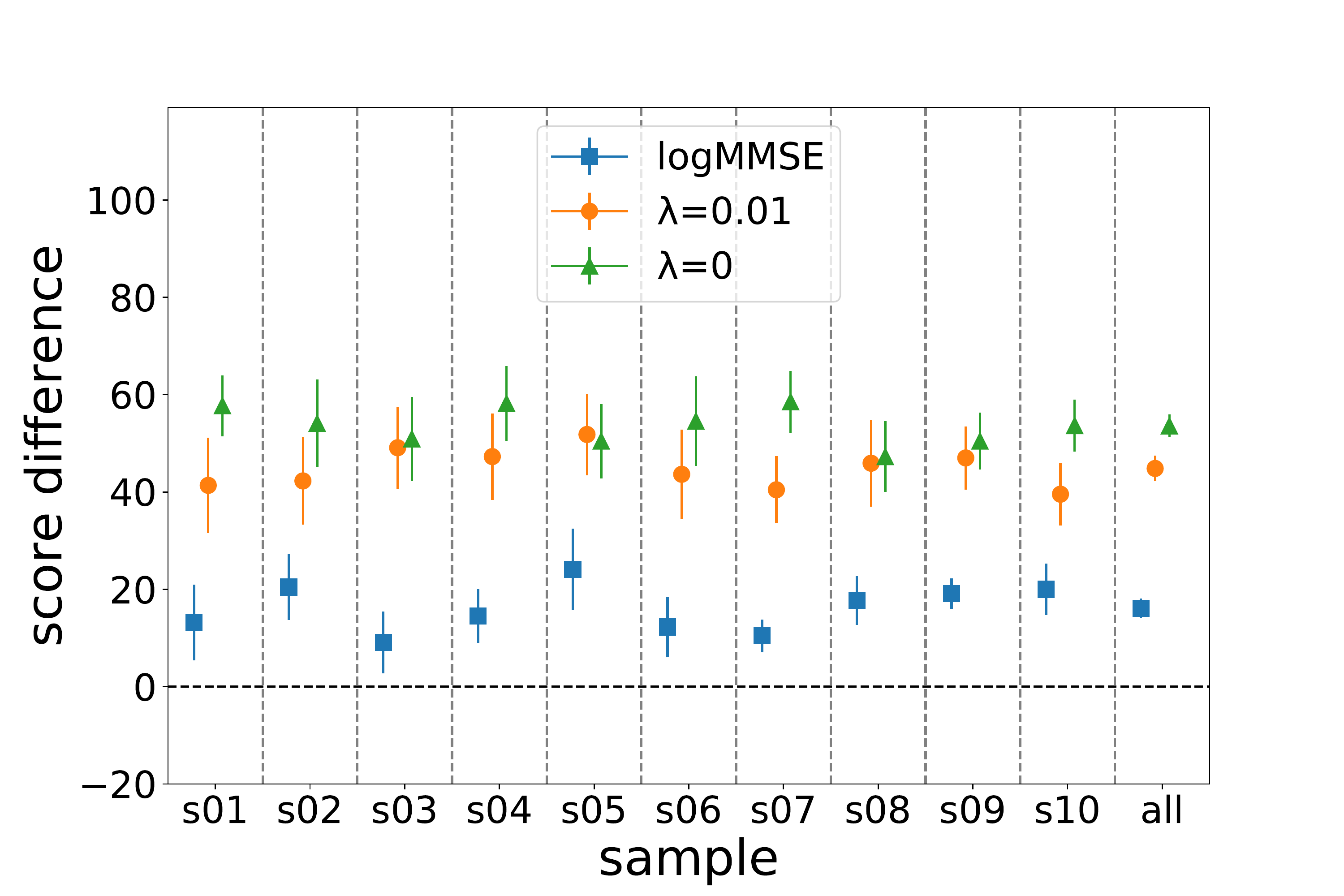}
\caption{Average score differences for the historical recordings dataset, relative to the original noisy sample.}
\label{fig:histrec_score_diffs}
\end{figure}
\begin{figure}
\centering
\includegraphics[width=0.95\linewidth]{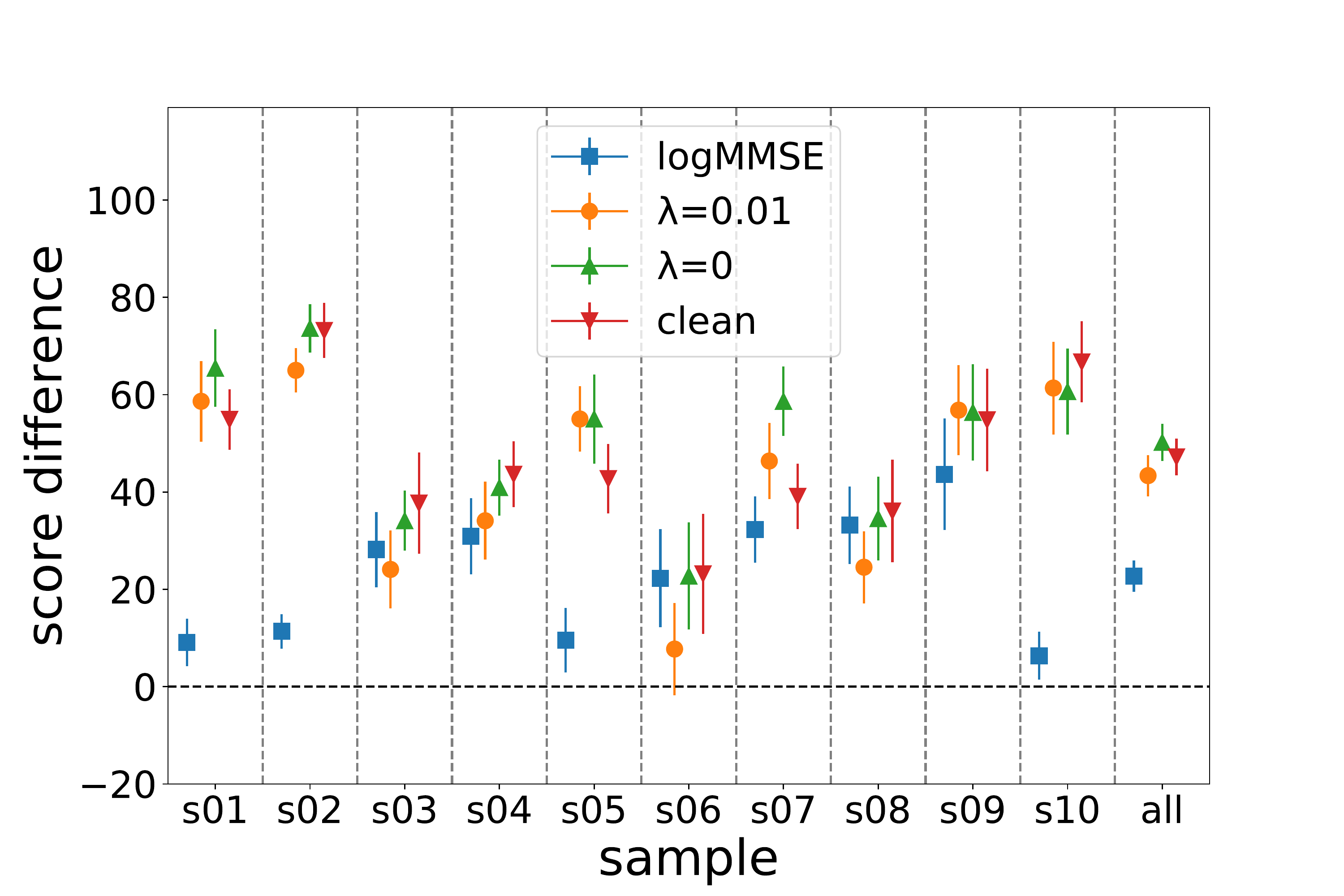}
\caption{Average score differences for the synthetic dataset, relative to the noisy sample.}
\label{fig:skyjam_score_diffs}
\end{figure}
We use the original noisy audio as the reference from which to compute score differences
for the historical recordings, and the synthetically noisified sample as the reference for the synthetic data. The results are shown in
Figure~\ref{fig:histrec_score_diffs}
for the historical recordings, and in
Figure~\ref{fig:skyjam_score_diffs}
for the synthetic dataset.
Error bars are 95\% confidence intervals, assuming a Gaussian distribution of the mean.
Both of our methods significantly improve the historical recordings, by around 50 points on average. In comparison, the logMMSE baseline only improves by an average of 16 points.
We also performed a Wilcoxon signed-rank test between our $\lambda$=0.01 and $\lambda$=0 models, to find that the difference is statistically significant (p-value $< 1.19\times 10^{-11}$).
On the synthetic data, again the $\lambda$=0 model outperforms the $\lambda$=0.01 variant,
with a p-value $< 2.13 \times 10^{-8}$.
On the other hand, there is no significant difference between the mean score differences of the $\lambda$=0 model and the clean sample (p-value = $0.097$).






\section{Conclusion}
\label{sec:conclusion}
%


We presented a learning-based method for automated denoising and applied it to restoration of noisy historical music recordings, matching a high quality bar:
Judged by human listeners on actual historical records, our method improves audio quality by a large margin and strongly outperforms existing approaches on a MUSHRA-like quality metric.
On artificially noisified music, it even attains a quality level that listeners found to be statistically indistinguishable from the ground truth.




\bibliography{references}

\begin{thebibliography}{10}
\providecommand{\url}[1]{#1}
\csname url@samestyle\endcsname
\providecommand{\newblock}{\relax}
\providecommand{\bibinfo}[2]{#2}
\providecommand{\BIBentrySTDinterwordspacing}{\spaceskip=0pt\relax}
\providecommand{\BIBentryALTinterwordstretchfactor}{4}
\providecommand{\BIBentryALTinterwordspacing}{\spaceskip=\fontdimen2\font plus
\BIBentryALTinterwordstretchfactor\fontdimen3\font minus
  \fontdimen4\font\relax}
\providecommand{\BIBforeignlanguage}[2]{{%
\expandafter\ifx\csname l@#1\endcsname\relax
\typeout{** WARNING: IEEEtran.bst: No hyphenation pattern has been}%
\typeout{** loaded for the language `#1'. Using the pattern for}%
\typeout{** the default language instead.}%
\else
\language=\csname l@#1\endcsname
\fi
#2}}
\providecommand{\BIBdecl}{\relax}
\BIBdecl

\bibitem{godsill1998}
S.~H. Godsill and P.~J. Rayner, \emph{Digital Audio Restoration: A Statistical
  Model Based Approach}, 1st~ed.\hskip 1em plus 0.5em minus 0.4em\relax Berlin,
  Heidelberg: Springer-Verlag, 1998.

\bibitem{berouti1979}
M.~{Berouti}, R.~{Schwartz}, and J.~{Makhoul}, ``Enhancement of speech
  corrupted by acoustic noise,'' in \emph{IEEE International Conference on
  Acoustics, Speech, and Signal Processing (ICASSP)}, vol.~4, 1979, pp.
  208--211.

\bibitem{kamath2002}
S.~Kamath and P.~Loizou, ``A multi-band spectral subtraction method for
  enhancing speech corrupted by colored noise,'' in \emph{IEEE International
  Conference on Acoustics, Speech, and Signal Processing (ICASSP)}, vol.~4, 05
  2002.

\bibitem{reddy2007}
A.~M. {Reddy} and B.~{Raj}, ``Soft mask methods for single-channel speaker
  separation,'' \emph{IEEE Transactions on Audio, Speech, and Language
  Processing}, vol.~15, no.~6, pp. 1766--1776, 2007.

\bibitem{grais2011}
E.~M. {Grais} and H.~{Erdogan}, ``Single channel speech music separation using
  nonnegative matrix factorization and spectral masks,'' in \emph{International
  Conference on Digital Signal Processing (DSP)}, 2011, pp. 1--6.

\bibitem{scalart1996}
P.~{Scalart} and J.~V. {Filho}, ``Speech enhancement based on a priori signal
  to noise estimation,'' in \emph{1996 IEEE International Conference on
  Acoustics, Speech, and Signal Processing Conference Proceedings}, vol.~2,
  1996, pp. 629--632.

\bibitem{attias2001}
H.~Attias, J.~C. Platt, A.~Acero, and L.~Deng, ``Speech denoising and
  dereverberation using probabilistic models,'' in \emph{Advances in Neural
  Information Processing Systems 13}, 2001, pp. 758--764.

\bibitem{loizou2005}
P.~C. {Loizou}, ``Speech enhancement based on perceptually motivated bayesian
  estimators of the magnitude spectrum,'' \emph{IEEE Transactions on Speech and
  Audio Processing}, vol.~13, no.~5, pp. 857--869, 2005.

\bibitem{denoising-musical-audio-2008}
C.~{Fevotte}, B.~{Torresani}, L.~{Daudet}, and S.~J. {Godsill}, ``Sparse linear
  regression with structured priors and application to denoising of musical
  audio,'' \emph{IEEE Transactions on Audio, Speech, and Language Processing},
  vol.~16, no.~1, pp. 174--185, 2008.

\bibitem{DengSESZFO20}
\BIBentryALTinterwordspacing
J.~Deng, B.~W. Schuller, F.~Eyben, D.~Schuller, Z.~Zhang, H.~Francois, and
  E.~Oh, ``Exploiting time-frequency patterns with {LSTM-RNNs} for low-bitrate
  audio restoration,'' \emph{Neural Computing and Applications}, vol.~32,
  no.~4, pp. 1095--1107, 2020. [Online]. Available:
  \url{https://doi.org/10.1007/s00521-019-04158-0}
\BIBentrySTDinterwordspacing

\bibitem{greek-folk-music-denoise-2014}
N.~{Bassiou}, C.~{Kotropoulos}, and I.~{Pitas}, ``Greek folk music denoising
  under a symmetric \textalpha-stable noise assumption,'' in \emph{10th
  International Conference on Heterogeneous Networking for Quality,
  Reliability, Security and Robustness}, 2014, pp. 18--23.

\bibitem{denoising-phonogram-2015}
V.~{Mach}, ``Denoising phonogram cylinders recordings using structured
  sparsity,'' in \emph{2015 7th International Congress on Ultra Modern
  Telecommunications and Control Systems and Workshops (ICUMT)}, 2015, pp.
  314--319.

\bibitem{inpainting-long-audio-segments-2018}
N.~Perraudin, N.~Holighaus, P.~Majdak, and P.~Balazs, ``Inpainting of long
  audio segments with similarity graphs,'' \emph{IEEE/ACM Transactions on
  Audio, Speech, and Language Processing}, vol.~PP, pp. 1--1, 02 2018.

\bibitem{audio-superresolution-2017}
V.~Kuleshov, S.~Z. Enam, and S.~Ermon, ``Audio super resolution using neural
  networks,'' in \emph{5th International Conference on Learning Representations
  (ICLR) 2017, Workshop Track, Toulon, France}, 2017.

\bibitem{temporal-film-2019}
S.~Birnbaum, V.~Kuleshov, Z.~Enam, P.~Koh, and S.~Ermon, ``Temporal film:
  Capturing long-range sequence dependencies with feature-wise modulations,''
  in \emph{Proc. 33rd Annual Conference on Neural Information Processing
  Systems (NeurIPS 2019)}, 2019.

\bibitem{deep-network-priors-2019}
\BIBentryALTinterwordspacing
M.~Michelashvili and L.~Wolf, ``Audio denoising with deep network priors,''
  \emph{CoRR}, vol. abs/1904.07612, 2019. [Online]. Available:
  \url{http://arxiv.org/abs/1904.07612}
\BIBentrySTDinterwordspacing

\bibitem{germain2018senet}
F.~G. Germain, Q.~Chen, and V.~Koltun, ``Speech denoising with deep feature
  losses,'' 2018.

\bibitem{segan-icassp-2017}
S.~Pascual, A.~Bonafonte, and J.~Serr\`{a}, ``{SEGAN}: Speech enhancement
  generative adversarial network,'' in \emph{Interspeech 2017, 18th Annual
  Conference of the International Speech Communication Association}, 08 2017,
  pp. 3642--3646.

\bibitem{rethage2018}
D.~{Rethage}, J.~{Pons}, and X.~{Serra}, ``A wavenet for speech denoising,'' in
  \emph{IEEE International Conference on Acoustics, Speech and Signal
  Processing (ICASSP)}, 2018, pp. 5069--5073.

\bibitem{donahue2017}
\BIBentryALTinterwordspacing
C.~Donahue, B.~Li, and R.~Prabhavalkar, ``Exploring speech enhancement with
  generative adversarial networks for robust speech recognition,'' \emph{CoRR},
  vol. abs/1711.05747, 2017. [Online]. Available:
  \url{http://arxiv.org/abs/1711.05747}
\BIBentrySTDinterwordspacing

\bibitem{audio-codec-icassp2020}
A.~Biswas and D.~Jia, ``Audio codec enhancement with generative adversarial
  networks,'' in \emph{IEEE International Conference on Acoustics, Speech, and
  Signal Processing (ICASSP)}, 2020.

\bibitem{mushra-2015}
\BIBentryALTinterwordspacing
``Method for the subjective assessment of intermediate quality levels of coding
  systems {ITU-Recommendation BS.1534-3},'' 2015. [Online]. Available:
  \url{www.itu.int/rec/R-REC-BS.1534}
\BIBentrySTDinterwordspacing

\bibitem{unet}
O.~Ronneberger, P.~Fischer, and T.~Brox, ``U-net: Convolutional networks for
  biomedical image segmentation,'' in \emph{Medical Image Computing and
  Computer-Assisted Intervention -- MICCAI 2015}, N.~Navab, J.~Hornegger, W.~M.
  Wells, and A.~F. Frangi, Eds.\hskip 1em plus 0.5em minus 0.4em\relax Cham:
  Springer International Publishing, 2015, pp. 234--241.

\bibitem{odena2016deconvolution}
\BIBentryALTinterwordspacing
A.~Odena, V.~Dumoulin, and C.~Olah, ``Deconvolution and checkerboard
  artifacts,'' \emph{Distill}, 2016. [Online]. Available:
  \url{http://distill.pub/2016/deconv-checkerboard}
\BIBentrySTDinterwordspacing

\bibitem{kumar2019}
K.~Kumar, R.~Kumar, T.~de~Boissiere, L.~Gestin, W.~Z. Teoh, J.~Sotelo,
  A.~de~Brebisson, Y.~Bengio, and A.~Courville, ``{MelGAN}: Generative
  adversarial networks for conditional waveform synthesis,'' 2019.

\bibitem{salimans2016}
T.~Salimans and D.~P. Kingma, ``Weight normalization: A simple
  reparameterization to accelerate training of deep neural networks,'' in
  \emph{Advances in Neural Information Processing Systems 29}, D.~D. Lee,
  M.~Sugiyama, U.~V. Luxburg, I.~Guyon, and R.~Garnett, Eds.\hskip 1em plus
  0.5em minus 0.4em\relax Curran Associates, Inc., 2016, pp. 901--909.

\bibitem{clevert2016}
S.~H. Djork-Arn\'{e}~Clevert, Thomas~Unterthiner, ``Fast and accurate deep
  network learning by exponential linear units (elus),'' in \emph{ICLR}, 2016.

\bibitem{ba2016}
J.~L. Ba, J.~R. Kiros, and G.~E. Hinton, ``Layer normalization,'' \emph{arXiv
  preprint arXiv:1607.06450}, 2016.

\bibitem{maas2013}
A.~L. Maas, A.~Y. Hannun, and A.~Y. Ng, ``Rectifier nonlinearities improve
  neural network acoustic models,'' in \emph{ICML Workshop on Deep Learning for
  Audio, Speech and Language Processing}, 2013.

\bibitem{abdulatif2019aegan}
S.~Abdulatif, K.~Armanious, K.~Guirguis, J.~T. Sajeev, and B.~Yang, ``Aegan:
  Time-frequency speech denoising via generative adversarial networks,'' 2019.

\bibitem{karras2018progressive-gan}
\BIBentryALTinterwordspacing
T.~Karras, T.~Aila, S.~Laine, and J.~Lehtinen, ``Progressive growing of gans
  for improved quality, stability, and variation,'' in \emph{6th International
  Conference on Learning Representations, {ICLR} 2018, Vancouver, BC, Canada,
  April 30 - May 3, 2018, Conference Track Proceedings}.\hskip 1em plus 0.5em
  minus 0.4em\relax OpenReview.net, 2018. [Online]. Available:
  \url{https://openreview.net/forum?id=Hk99zCeAb}
\BIBentrySTDinterwordspacing

\bibitem{armanious2018medgan}
\BIBentryALTinterwordspacing
K.~Armanious, C.~Yang, M.~Fischer, T.~K{\"{u}}stner, K.~Nikolaou, S.~Gatidis,
  and B.~Yang, ``Medgan: Medical image translation using {GANs},'' \emph{CoRR},
  vol. abs/1806.06397, 2018. [Online]. Available:
  \url{http://arxiv.org/abs/1806.06397}
\BIBentrySTDinterwordspacing

\bibitem{adam-optimizer}
D.~Kingma and J.~Ba, ``Adam: A method for stochastic optimization,''
  \emph{International Conference on Learning Representations}, 12 2014.

\bibitem{public-domain-project-histrec}
\BIBentryALTinterwordspacing
``Public domain project,'' \url{http://pool.publicdomainproject.org}, [Online;
  accessed February-2020]. [Online]. Available:
  \url{http://pool.publicdomainproject.org}
\BIBentrySTDinterwordspacing

\bibitem{hershey2017vggish}
\BIBentryALTinterwordspacing
S.~Hershey, S.~Chaudhuri, D.~P.~W. Ellis, J.~F. Gemmeke, A.~Jansen, C.~Moore,
  M.~Plakal, D.~Platt, R.~A. Saurous, B.~Seybold, M.~Slaney, R.~Weiss, and
  K.~Wilson, ``{CNN} architectures for large-scale audio classification,'' in
  \emph{International Conference on Acoustics, Speech and Signal Processing
  (ICASSP)}, 2017. [Online]. Available: \url{https://arxiv.org/abs/1609.09430}
\BIBentrySTDinterwordspacing

\bibitem{fad-interspeech-2019}
\BIBentryALTinterwordspacing
K.~Kilgour, M.~Zuluaga, D.~Roblek, and M.~Sharifi, ``Fr{\'{e}}chet audio
  distance: {A} reference-free metric for evaluating music enhancement
  algorithms,'' in \emph{Interspeech 2019, 20th Annual Conference of the
  International Speech Communication Association, Graz, Austria, 15-19
  September 2019}, G.~Kubin and Z.~Kacic, Eds.\hskip 1em plus 0.5em minus
  0.4em\relax {ISCA}, 2019, pp. 2350--2354. [Online]. Available:
  \url{https://doi.org/10.21437/Interspeech.2019-2219}
\BIBentrySTDinterwordspacing

\bibitem{ephraim1985}
Y.~{Ephraim} and D.~{Malah}, ``Speech enhancement using a minimum mean-square
  error log-spectral amplitude estimator,'' \emph{IEEE Transactions on
  Acoustics, Speech, and Signal Processing}, vol.~33, no.~2, pp. 443--445,
  1985.

\bibitem{scipy2020}
P.~{Virtanen}, R.~{Gommers}, T.~E. {Oliphant}, M.~{Haberland}, T.~{Reddy},
  D.~{Cournapeau}, E.~{Burovski}, P.~{Peterson}, W.~{Weckesser}, J.~{Bright},
  S.~J. {van der Walt}, M.~{Brett}, J.~{Wilson}, K.~{Jarrod Millman},
  N.~{Mayorov}, A.~R.~J. {Nelson}, E.~{Jones}, R.~{Kern}, E.~{Larson},
  C.~{Carey}, {\.I}.~{Polat}, Y.~{Feng}, E.~W. {Moore}, J.~{Vand erPlas},
  D.~{Laxalde}, J.~{Perktold}, R.~{Cimrman}, I.~{Henriksen}, E.~A. {Quintero},
  C.~R. {Harris}, A.~M. {Archibald}, A.~H. {Ribeiro}, F.~{Pedregosa}, P.~{van
  Mulbregt}, and S.~.~. {Contributors}, ``{SciPy 1.0: Fundamental Algorithms
  for Scientific Computing in Python},'' \emph{Nature Methods}, vol.~17, pp.
  261--272, 2020.

\end{thebibliography}

\end{document}